\documentclass[twocolumn]{revtex4}
\usepackage[english]{babel}
\usepackage[ansinew]{inputenc}
\usepackage[T1]{fontenc}

\usepackage{graphicx}
\usepackage{amsmath,bm}
\usepackage{amssymb}

\usepackage{color}

\newcommand{\abs}[1]{\lvert #1 \rvert}
\newcommand{\tuborg}[1]{\left\{ #1 \right\}}
\newcommand{\af}[1]{\left( #1 \right)}
\newcommand{\kantpar}[1]{\left[ #1 \right]}
\newcommand{\FigRef}[1]{{\figurename~\ref{#1}}}

\newcommand{\WedgeProj}{\Op{\mathcal W}}
\newcommand{\defi}{\equiv}

\newcommand{\Op}[1]{\hat{#1}}
\newcommand{\HC}{\dagger}

\newcommand{\aOp}{{\Op a}^\HC}
\newcommand{\aNed}{\Op a}
\newcommand{\Identity}{\Op 1}

\newcommand{\HermitianConjugate}{\mathrm{H.c.}}

\newcommand{\imag}{\mathrm{i}}
\newcommand{\Exp}[1]{\mathrm{e}^{#1}}

\newcommand{\Hilbert}[2]{\mathcal{H}^{#1}_{#2}}
\newcommand{\FockBasisStates}{\mathcal{B}}
\newcommand{\Equivalence}{\mathcal{E}}

\newcommand{\Occupation}{\mathcal{O}}
\newcommand{\Tauk}[1][]{\Op\tau_{#1}}

\newcommand{\EqBasis}[2]{{\Phi^{(#1)}_{#2}}}
\newcommand{\EState}[2]{{\Psi^{#1}_{#2}}}

\newcommand{\AState}[2]{{\alpha^{#1}_{#2}}}
\newcommand{\Obasis}[2][]{{\Omega^{#1}_{#2}}}
\newcommand{\Order}[1]{{\Lambda_{#1}}}
\newcommand{\TotNum}[2]{{\Gamma^{#1}_{#2}}}

\newcommand{\Translation}{\Op{\mathcal{T}}}

\newcommand{\leqN}[1]{\preceq_{#1}}

\DeclareMathAlphabet{\mathboring}{OT1}{cmss}{bx}{it}

\newcommand{\OBDM}{\rho}

\newcommand{\GroundState}[2]{\chi^{#1}_{#2}}
\newcommand{\FirstExcited}[2]{\sigma^{#1}_{#2}}
\newcommand{\SecondExcited}[2]{\kappa^{#1}_{#2}}

\def\ket#1{\mathinner{|{#1}\rangle}}
\def\braket#1{\mathinner{\langle{#1}\rangle}}

\let\protect\relax
{\catcode`\|=\active
  \xdef\Braket{\protect\expandafter\noexpand\csname Braket \endcsname}
  \expandafter\gdef\csname Braket \endcsname#1{\begingroup
     \ifx\SavedDoubleVert\relax
       \let\SavedDoubleVert\|\let\|\BraDoubleVert
     \fi
     \mathcode`\|32768\let|\BraVert
     \left\langle{#1}\right\rangle\endgroup}
}
\def\BraVert{\@ifnextchar|{\|\@gobble}
     {\egroup\,\mid@vertical\,\bgroup}}
\def\BraDoubleVert{\egroup\,\mid@dblvertical\,\bgroup}
\let\SavedDoubleVert\relax

{\catcode`\|=\active
  \xdef\set{\protect\expandafter\noexpand\csname set \endcsname}
  \expandafter\gdef\csname set \endcsname#1{\mathinner
        {\lbrace\,{\mathcode`\|32768\let|\midvert #1}\,\rbrace}}
  \xdef\Set{\protect\expandafter\noexpand\csname Set \endcsname}
  \expandafter\gdef\csname Set \endcsname#1{\left\{%
     \ifx\SavedDoubleVert\relax \let\SavedDoubleVert\|\fi
     \:{\let\|\SetDoubleVert
     \mathcode`\|32768\let|\SetVert
     #1}\:\right\}}
}
\def\midvert{\egroup\mid\bgroup}
\def\SetVert{\@ifnextchar|{\|\@gobble}
    {\egroup\;\mid@vertical\;\bgroup}}
\def\SetDoubleVert{\egroup\;\mid@dblvertical\;\bgroup}

\begingroup
 \edef\@tempa{\meaning\middle}
 \edef\@tempb{\string\middle}
\expandafter \endgroup \ifx\@tempa\@tempb
 \def\mid@vertical{\middle|}
 \def\mid@dblvertical{\middle\SavedDoubleVert}
\else
 \def\mid@vertical{\mskip1mu\vrule\mskip1mu}
 \def\mid@dblvertical{\mskip1mu\vrule\mskip2.5mu\vrule\mskip1mu}
\fi

\begin{document}

\title{Relative and center-of-mass motion in the attractive Bose-Hubbard model}

\author{Ole S\o e S\o rensen}
\affiliation{Lundbeck Foundation Theoretical Center for Quantum System Research, Department of Physics and Astronomy, Aarhus University, DK-8000 Aarhus C, Denmark}

\author{S\o ren Gammelmark}
\affiliation{Lundbeck Foundation Theoretical Center for Quantum System Research,  Department of Physics and Astronomy, Aarhus University, DK-8000 Aarhus C, Denmark}
\author{Klaus M\o lmer}
\affiliation{Lundbeck Foundation Theoretical Center for Quantum System Research,  Department of Physics and Astronomy, Aarhus University, DK-8000 Aarhus C, Denmark}

\begin{abstract}

We present first-principle numerical calculations for few particle solutions of the attractive Bose-Hubbard model with periodic boundary conditions.
We show that the low-energy many-body states found by numerical diagonalization can be written as translational superposition states of compact composite systems of particles. These compact states break the translational symmetry of the problem and their center-of-mass and internal excitations offer simple explanations of the energy spectrum of the full model.
\end{abstract}
\maketitle

\section{Introduction}

Interacting many-body quantum systems present theoretical challenges in all branches of physics and chemistry, and a wide range of methods have been developed to provide efficient computational methods and to gain insight in their detailed structure and dynamics. Separation of different physical degrees of freedom is a widely used approach, e.g., in chemistry, where the Born-Oppenheimer approximation treats the electronic motion in the presence of classically fixed nuclei, and subsequently the resulting energy levels together with the direct nuclear interaction constitute potential energy surfaces for the quantum motion of the nuclei \cite{Herzberg}. The nuclear motion, in turn, can be separated into vibrational motion around minimum energy configurations, and quantum translational and rotational motion of the entire molecule. The latter reflect the translational and rotational invariance of the Hamiltonian and establish total momentum and angular momentum as conserved quantities with associated good quantum numbers.
\begin{figure}[b]
  \centering  
  \includegraphics{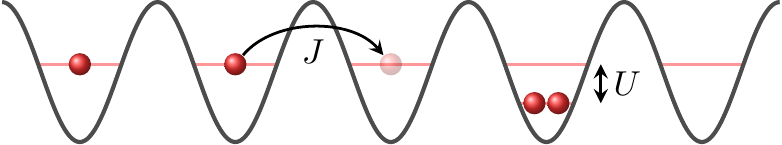}  
  \includegraphics[width = \columnwidth]{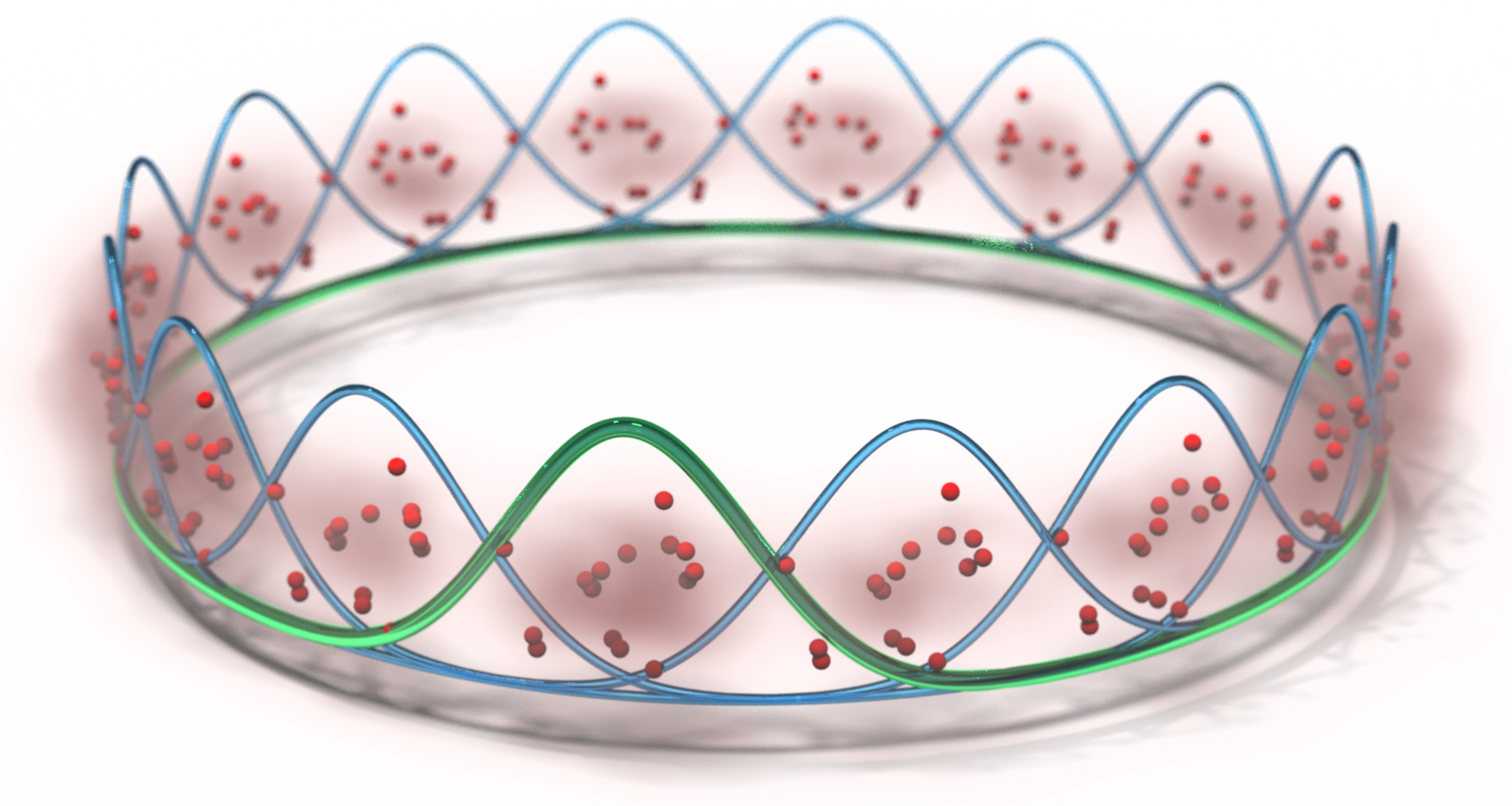}  
  \caption{
    The Bose-Hubbard model describing atoms populating single particle states localized at lattice potential minima. Upper panel: The atoms experience tunneling between sites, while atoms on the same site interact with interaction strength $U<0$. Lower panel: Many-body quantum state in a finite lattice potential with periodic boundary conditions. The translated copies of a classical configuration of atoms depicted as red balls illustrate how the full state is a superposition of translated many-body quantum states. Similarly, the green curve shows the atomic density in a localized state, while the density in the translated replica are indicated by the blue curves.}
  \label{fig:SchematicConcept}
\end{figure}

A similar separation of the interaction energy associated with the relative motion and with the translation and rotation of the composite object of particles
is used for the state of atomic nuclei \cite{Bohr}, and it has also been applied in the interpretation of excited state spectra of atoms with spatially (anti-)correlated electrons  \cite{Taulbjerg,Rost,lithium}. The validity of such a separation in relative motion of individual particles and  collective motion of the bound composite object is usually explained with reference to a separation in energy scales which suppresses the coupling between the different degrees of freedom and justifies an adiabatic separation, e.g., of the slow nuclear and fast electronic motion in molecules.

In ultra-cold atom physics it is possible to study the case of binding due to very weak and
tunable
interactions, and we may probe the transition between a collection of independent particles and a single massive object held together by the attractive interactions. Attractive bosons may thus form a so-called bright-soliton, which in mean field theory \cite{Shabat-Inverse-Scattering, Garcia, Carr} is described by a localized solution to the non-linear Schrödinger Equation. The choice of a particular localization breaks the translational symmetry of the system in free space, but rather than restoring translational symmetry by diluting the atoms in a uniform mean field wave function, a quantum superposition state of translationally displaced copies of the soliton wave function may form a better Ansatz for the many-body quantum state.

A similar situation occurring for atoms in a trap was analyzed by Pethick and Pitaevskii \cite{PethickPitaevskii}, see also \cite{Gajda, Zinner, Yamada}, where again a compact bright soliton like object may form, and where this object may exercise quantum motion described by the center-of-mass ground state wave function in the trap. If the quantum mechanical excursions of the center-of-mass motion are comparable to or larger than the soliton width,
the state is poorly described by a single mean field wave function, and as a consequence the high degree of order in the system is not reflected by a macroscopic population of a single particle state. 

In \cite{Zin, Kanamoto} mean-field and Bogoliubov theories are used to reveal how a system of bosons on a ring with periodic boundary conditions breaks the translational symmetry and undergoes a quantum phase transition towards a localized soliton-like state, and how critical fluctuations and localization may occur due to a weak symmetry-breaking perturbation or measurements on the system. Anderson localization of the mean field solution in an external disordered potential on length scales much larger than the soliton width is reported in \cite{Sacha}.

The separation between relative degrees of freedom and center-of-mass motion of composite, bound objects, does not rely on mean field theory. In \cite{WeissCastin} it is thus shown that the soliton-like object can scatter and form macroscopic superposition states of transmitted and reflected components, while a many body description is used to describe the decay of the soliton into fragmented condensates in \cite{Streltsov}. Two particle bound states on a lattice separate exactly in center-of-mass and relative coordinates  \cite{Winkler, Manuel, Nygaard}, and in \cite{CastinHerzog}, it is shown how the ground state solution to the Lieb-Liniger model of an attractive Bose gas in one dimension can be precisely represented as a translational invariant state with a definite dependence on the relative coordinates between the particles following a Bethe Ansatz with exponentially decreasing terms. 

In this article, we will analyze the attractive Bose-Hubbard model on a finite lattice with periodic boundary conditions as illustrated in \FigRef{fig:SchematicConcept}. We will treat the case of few atoms and few sites, which permits exact numerical diagonalization of the problem, and we will subsequently analyze the results, and show how the discrete translational symmetry of the problem gives rise to states, which can be represented as superpositions of translated versions of compact symmetry breaking states. We will study both the ground state and excited states of the system, and we will identify how the low-energy part of the energy spectrum can be understood in terms of the center-of-mass and relative motion of the atoms.

In Sec.~\ref{sec:TheSystem}, we introduce the system and identify the symmetries of the problem which enable a reduction of the full many-body problem to a numerically tractable form. In Sec.~\ref{sec:NumericalResults}, we present numerically determined energy spectra and correlation functions in support of our separation of the problem. In Sec.~\ref{sec:Decoupling}, we show that our states can be represented by the desired Ansatz. In Secs.~\ref{sec:LocalizedStates} and~\ref{sec:EnergyOfTranslationalMotion}, we show how the underlying compact quantum states are related to and can explain features in the computed energy spectrum for the model. We conclude in Sec.~\ref{sec:Conclusion}

\section{The system and its symmetries}
\label{sec:TheSystem}
We consider a one-dimensional lattice with $M$ sites numbered $0,1 , 2 , \ldots , M-1$, and we restrict our analysis to the case where atoms can access only one single-particle state $\ket{\phi_j}$ at each site $j$. A system of identical bosons on such a lattice is most conveniently treated in second quantization, and in the tight-binding approximation, the system is well-described by the Bose-Hubbard Hamiltonian
\begin{align} \label{BH1}
  \Op H &= - J \sum_j \af{ \aOp_{j+1} \aNed_j + \HermitianConjugate}
  + \frac{U}{2} \sum_j \Op n_j (\Op n_j - 1) ,
\end{align}
where $J$ and $U$ are the tunneling and the on-site interaction strengths, respectively. In the present paper we consider attractive interactions, $U \leq 0$. The operator $\aOp_j$ creates a particle in the single particle state $\ket{\phi_j}$, and $\Op n_j \defi \aOp_j \aNed_j$ is the corresponding number operator on the $j$th site.
Periodic boundary conditions are ensured by letting all site indices be implicitly understood as modulo $M$.

$\Op H$ conserves the number of particles, and we will perform our calculations within fixed Hilbert spaces, $\Hilbert{M}{N}$, of states with $N$ particles on $M$ sites, conveniently spanned by the basis $\FockBasisStates_N^M$ of multi-mode Fock states $\ket{n_0 , \ldots , n_{M-1}}$ where $n_0 + \cdots + n_{M-1} = N$. There are
\begin{align}
  \TotNum{M}{N} &\defi \binom{M + N - 1}{N}
  \label{eq:TotNum}
\end{align}
elements in the set of such basis vectors, and in \appendixname~\ref{sec:Ordering}, we discuss an ordering of these basis states which makes the action of the Hamiltonian particularly easy to calculate.

The unitary many-body translation operator, $\Translation$, acts on the many-body state by moving all particles one site ``to the right'':
\begin{align}
  \Translation \ket{n_0 , \ldots , n_{M-1}}
  &\defi \ket{n_{M-1} , n_0 , n_1, \ldots , n_{M-2}},
\end{align}
and when we apply it $M$ times on any of the Fock basis states, we return to the original state, i.e.
\begin{align}
  \Translation^M = \Identity .
  \label{eq:TCyclic}
\end{align}
\begin{figure*}[tb]
  \centering
  \includegraphics[width=\textwidth]{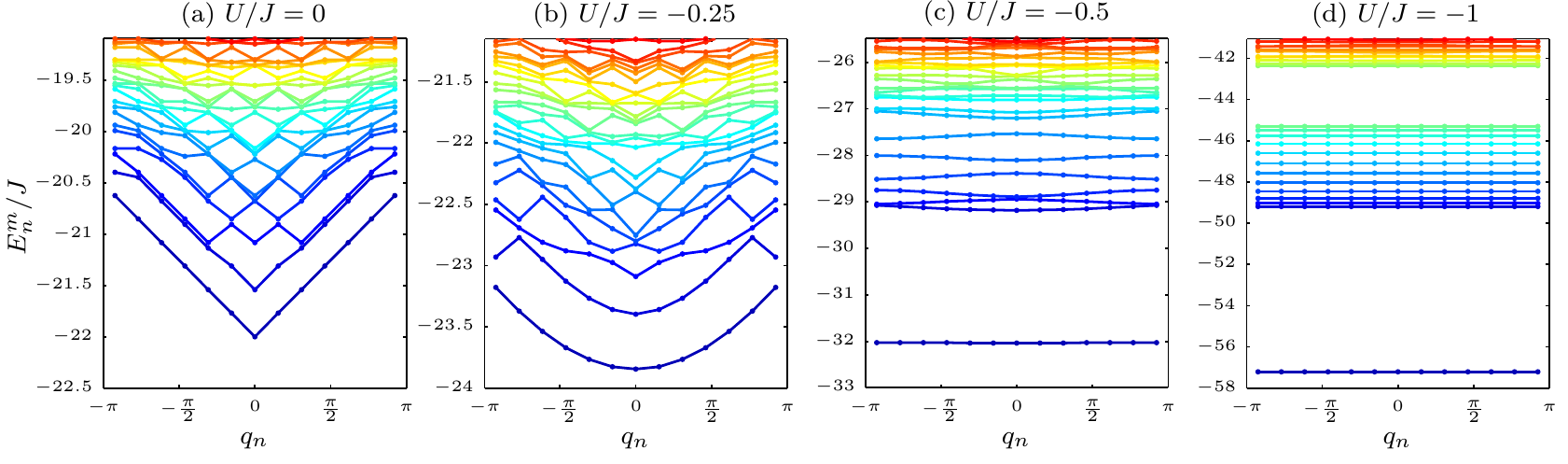}
  \caption{(Color online) Energy spectra for 11 particles on an 13 site lattice are shown as functions of the quasi-momentum $q_n$ for (a) $U/J=0$, (b) $U/J=-0.25$, (c) $U/J=-0.5$, and (d) $U/J=-1$. The colored solid lines connect the energies computed at the discrete allowed values of $q_n$.}
  \label{fig:DispersionRelation}
\end{figure*}
The action of the translation operator can also be specified by the relations
\begin{align}
  \Translation \aOp_j &= \aOp_{j+1} \Translation
  , &
  \Translation \aNed_j &= \aNed_{j+1} \Translation ,
\end{align}
and their hermitian conjugates. It follows directly that the Bose-Hubbard Hamiltonian commutes with $\Translation$ and thus
simultaneous eigenstates exist for the two operators. Since the relation~(\ref{eq:TCyclic}) dictates that the $\Translation$-eigenvalues are $M$th roots of unity, we can label those eigenstates $\ket{\EState{m}{n}}$ as
\begin{gather}
  \begin{aligned}
    \Op H \ket{\EState{m}{n}} &= E^m_n \ket{\EState{m}{n}}
    , \\
    \Translation \ket{\EState{m}{n}} &= \Exp{-\imag q_n} \ket{\EState{m}{n}} .
  \end{aligned}
  \label{eq:EnergyEigenstates}
\end{gather}
In the limit of a large number of sites, the energies $E^m_n$ form bands labeled by the excitation number $m = 1,2,\ldots$ and by the quasi-momentum
\begin{align}
  q_n &\defi \frac{2 \pi n}{M} \in  ]- \pi ; \pi],
\end{align}
which, acquires $M$ discrete values in a finite lattice.

Finally, the system is invariant under parity symmetry, which ensures that
$E^m_n = E^m_{-n}$
and which imposes constraints on the real and imaginary parts of our expansion coefficients that are useful in the following.

We can divide the basis $\FockBasisStates_N^M$ into $P$ disjoint equivalence classes of basis vectors
\begin{align}
  \FockBasisStates_N^M &= \Equivalence^{(1)} \sqcup \Equivalence^{(2)} \sqcup \cdots \sqcup \Equivalence^{(P)}
  \label{eq:DefinitionOfEquivalenceClasses}
\end{align}
where $\ket{\Obasis{a}}$ and $\ket{\Obasis{b}}$ lie in the same equivalence class $\Equivalence^{(j)}$ iff $\ket{\Obasis{b}} = \Translation^k \ket{\Obasis{a}}$ for some $k$. Every state $\ket\Psi$  can be expanded uniquely as
\begin{align}
  \ket{\Psi} &= \sum_{j = 1}^P \sum_{k = 0}^{\abs{\Equivalence^{(j)}}-1} c^{(j)}_k \ket{\EqBasis{j}{k}},
  \label{eq:EquivalenceClassExpansion}
\end{align}
where we have renamed the basis states such that
\begin{align}
  \Equivalence^{(j)}
  &= \tuborg{\ket{\EqBasis{j}{0}} , \ket{\EqBasis{j}{1}} , \ldots , \ket{\EqBasis{j}{\abs{\Equivalence^{(j)}}-1}}},
  \label{eq:StatesOfAnEquivalenceClass}
\end{align}
$\abs{\Equivalence^{(j)}}$ being the number of elements in $\Equivalence^{(j)}$, and $\ket{\EqBasis{j}{k}} = \Translation^{k}\ket{\EqBasis{j}{0}}$.

If $\ket{\Psi}$ is a translation eigenstate with quasi-momentum $q_n$, then for any two elements $\ket{\EqBasis{j}{a}}$ and $\ket{\EqBasis{j}{b}}$ from $\Equivalence^{(j)}$,
\begin{align*}
  \braket{\EqBasis{j}{a} | \Psi}
  &= \braket{\EqBasis{j}{b} | \Translation^{b - a} | \Psi}
  = \Exp{\imag (a-b) q_n}\braket{\EqBasis{j}{b} | \Psi},
\end{align*}
and the expansion~(\ref{eq:EquivalenceClassExpansion}) takes the simpler form
\begin{align}
  \ket{\Psi}
  &= \sum_{j = 1}^P C_j  \sum_{k = 0}^{\abs{\Equivalence^{(j)}}-1} \Exp{\imag k q_n} \ket{\EqBasis{j}{k}}.
  \label{eq:ExpansionOfTranslationEigenstates}
\end{align}
where $C_j$  accounts for the overlap between $\ket\Psi$ and states from $\Equivalence^{(j)}$.

Since $\Op H$ and $\Translation$ commute, $\Op H$ is block diagonal in a basis of translation eigenstates.
Inspired by the expansion~(\ref{eq:ExpansionOfTranslationEigenstates}) we can choose such a basis as
\begin{align}
  \ket{\xi_n^{(j)}}
  &\defi \sqrt{\frac{\abs{\Equivalence^{(j)}}}{M}} \sum_{k = 0}^{\abs{\Equivalence^{(j)}}-1} \Exp{-2 \pi \imag k n/M} \ket{\EqBasis{j}{k}}
\end{align}
with $n = 0 , \ldots , M-1$ and $j = 1,\ldots ,P$. These states are orthonormal in both indices and the matrix representation of $\Op H$ is block-diagonal. The problem thus reduces to the identification of eigenvalues and vectors of $M$ matrices of dimension $\TotNum{M}{N}/M$ rather than of one matrix of dimension $\TotNum{M}{N}$. 

\section{Numerical results}
\label{sec:NumericalResults}
Using the state vector representation described in the Appendix, and the simplification of the problem following from the above analysis, the system becomes amenable to standard numerical diagonalization routines, and the eigenstates  and -energies (\ref{eq:EnergyEigenstates}) of $\Op H$ can be found.

\subsection{Energy spectra}

In \FigRef{fig:DispersionRelation}, the energy spectra for 11 particles on a 13 site lattice are shown for different interaction strengths delineating the transition from the non-interacting regime to the strongly attractive regime. The colored solid lines are used to connect the energies at the discrete allowed values of $q_n$ in the figure.

In part (a) of the figure, $U/J = 0$ and the interaction term of $\Op H$ vanishes. Thus, the system can be solved by finding the single-particle eigenstates and -energies on the discrete lattice and forming $N$-particle product states. Weak interactions among the atoms are introduced in part (b) of the figure. They couple the product states, and as the interactions get stronger in parts (c) and (d) of the figure, the spectrum develops structures at different scales. These are the structures that we will associate with different modes of atomic motion: low energy translational motion of the entire composite object, slightly higher energy motion of atoms outside the bound composite object, and large energy gaps associated with the excitation of different numbers of atoms out of the bound composite object.

In the subsequent sections, we examine this separation hypothesis further by an analysis of the eigenstates found by our diagonalization, and we shall also quantitatively address the energy contributions from center-of-mass and individual particle motion.

Here, let us consider the coarse scale structures in \FigRef{fig:DispersionRelation}(d) 
The interaction energy  of $n$ particles on a single site is $\frac{U}{2}n(n-1)$, and this explains the coarse separation of the spectrum into bands of states around $\frac{U}{2}N(N-1)$, $\frac{U}{2}(N-1)(N-2)$, and $\frac{U}{2}(N-2)(N-3)$, which are translational superpositions of states where either all atoms occupy the localized condensate state, or one or two atoms are excited out of the condensate.
In the total absence of tunneling, $U/J \to -\infty$, each of these bands become degenerate, while in \FigRef{fig:DispersionRelation}(d), in the presence of both tunneling and strong attraction, the degeneracy is lifted within each quasi-momentum subspace, and the bands split into a series of almost flat sub-bands. The crossing structures observed at the weaker interaction strength in \FigRef{fig:DispersionRelation}(c), will be discussed below.

\subsection{Eigenstates}

\begin{figure}[tb]
  \centering
  \includegraphics[width=\columnwidth]{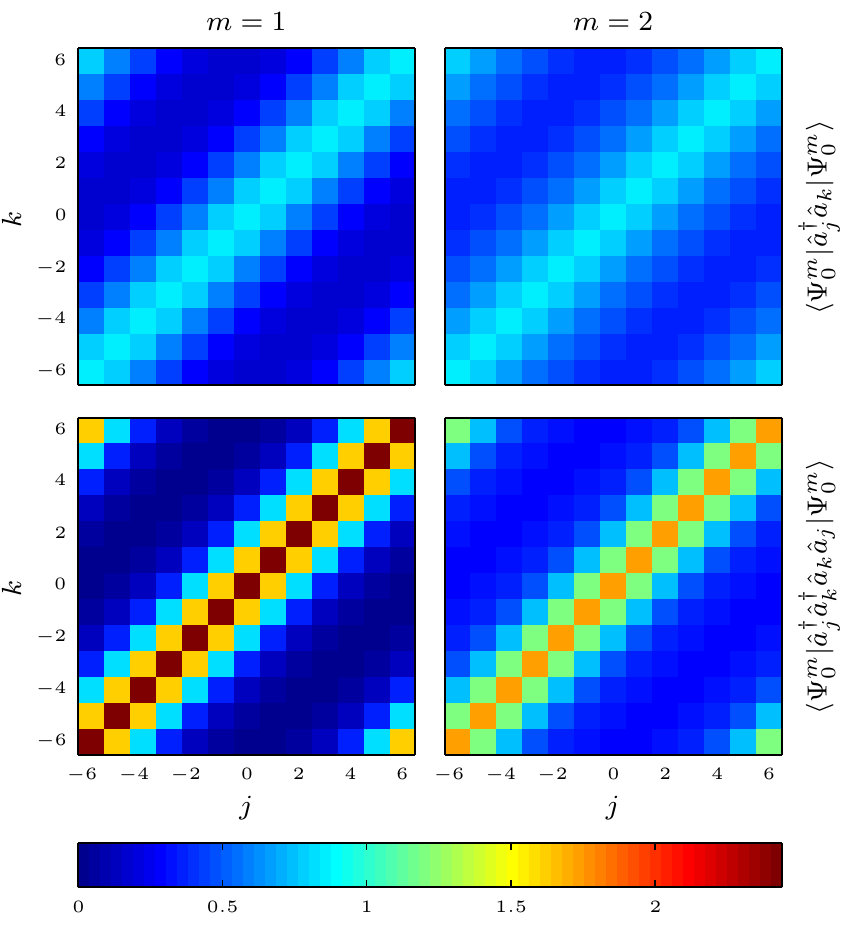}
  \caption{(Color online) First and second order coherence functions, $\braket{\EState{m}{n} | \aOp_j \aNed_k | \EState{m}{n}}$ and  $\braket{\EState{m}{n} | \aOp_j \aOp_k \aNed_k\aNed_j | \EState{m}{n}}$, for the two lowest many-body eigenstates with vanishing quasi-momentum for $U/J = -0.25$.}
  \label{fig:CorrelationFunctions}
\end{figure}

All the energy eigenstates found by our diagonalization are quasi-momentum eigenstates, and hence the particle density is uniformly distributed over all lattice sites. Usually, the structure within the many-body states is analyzed by correlation functions, and in \FigRef{fig:CorrelationFunctions}, we show the first order (amplitude) coherence function,  $\braket{\EState{m}{n} | \aOp_j \aNed_k | \EState{m}{n}}$, (the one-body density matrix), and the second order (density) coherence function, $\braket{\EState{m}{n} | \aOp_j \aOp_k \aNed_k\aNed_j | \EState{m}{n}}$. The functions are shown for the two lowest energy states with vanishing quasi-momentum. The upper plots reveal the constant density along the identical $j=k$ diagonal elements, and they show the vanishing of the first order coherence function for large separation. Diagonalization of the one-body density matrix, yields $M$ almost identical eigenvalues around $N/M$, indicating that the atoms are evenly distributed on $M$ single particle modes, and according to the Penrose-Onsager criterion, the system of bosons does not populate a condensate. The width of the diagonal bands in the figure indicate, however, that these single particle modes are localized on a small number of neighboring sites. This is similar to the observation  by Pethick and Pitaevskii \cite{PethickPitaevskii} that trapped attractive gases do  not condense according to the conventional criteria of Bose-Einstein condensation.

The lower plots show the density coherence functions with the intuitive interpretation as the probability of finding a particle at a given site, $k$, given that another particle has been seen at another site, $j$. The large values along the diagonal show that, while evenly distributed on average, the atoms are not independent, but cluster strongly around each other within a range comparable to the single-particle coherence range in the upper panels. This of course supports our hypothesis, that the atoms form a compact condensate wave function, like a bright solitonic wave in mean field theory, but in order to minimize its kinetic energy, the center-of-mass of this wave function should be completely delocalized over the lattice.

We note, that localized bright or dark solitons in mean field theory break the translational symmetry of the underlying equations, while Bogoliubov theory returns both quasi-particle excitation modes and two Goldstone modes associated with the $U(1)$ phase symmetry breaking and the breaking of translational symmetry. The latter Goldstone mode causes a growing position uncertainty of the whole mean field solution and shows explicitly that localized solitons are not stationary solutions \cite{Negretti, Dziarmaga}. Similarly, in our case, a localized many-body state would not be an energy eigenstate.
We do not have recourse to Bogoliubov theory, but in the following section, we shall see how the same conceptual idea works in our exact solutions of the Bose-Hubbard model.

\begin{figure}[tb]
  \centering
  \includegraphics[width=0.8\columnwidth]{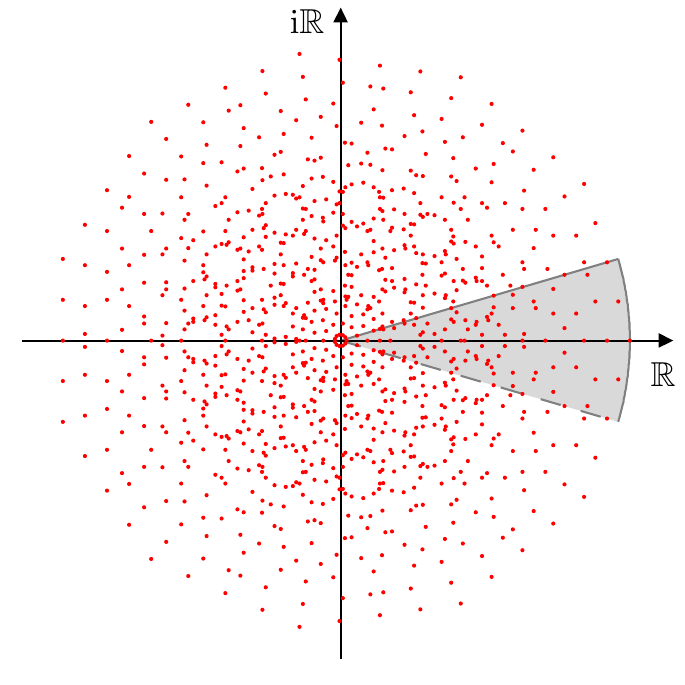}
  \caption{(Color online) Complex center-of-mass coordinates of all the Fock basis states for 11 modes and 4 particles. The gray wedge encapsulates states with a center of mass coordinate with a complex phase in the interval $\left]-\pi/M ; \pi/M \right]$.}
  \label{fig:Wedge}
\end{figure}

\section{Decoupling of the translational motion of eigenstates}
\label{sec:Decoupling}
The goal of this section is to identify localized, compact states $\ket{\AState{m}{n}}$ of the $N$ atoms, which enable us to write the translation eigenstates $\ket{\EState{m}{n}}$ as superposition states of the form,
\begin{align}
  \ket{\EState{m}{n}}
  &= C \sum_{k = 0}^{M-1} \Exp{\imag kq_n} \Translation^k \ket{\AState{m}{n}},
  \label{eq:TranslationSuperposition}
\end{align}
where $C$ is a normalization constant.
The lower part of \FigRef{fig:SchematicConcept} attempts to illustrate the idea behind an Ansatz of precisely this form, where the phase factors in~(\ref{eq:TranslationSuperposition}) lead to a translation eigenstate with quasi-momentum $q_n$.

\subsection{Superposition and projection}

We introduce Eq. (\ref{eq:TranslationSuperposition}) as a superposition of translated replica of the state $\ket{\AState{m}{n}}$, but we may also read the equation in a different manner by noticing that the self-adjoint operator
\begin{align}
  \Tauk[n]
  &\defi \frac{1}{M} \sum_{k = 0}^{M-1} \Exp{\imag kq_n} \Translation^k
  \label{eq:TauknDefinition}
\end{align}
is a projection operator, which projects any state in $\Hilbert{M}{N}$ onto the subspace of translation eigenstates with quasi-momentum $q_n$. The superposition~(\ref{eq:TranslationSuperposition}) of translated replica of a single localized state is thus, up to normalization, identical to the projection of that state on its $q_n$ quasi-momentum eigenstate component.

An immediate consequence of this is that given the target state $\ket{\EState{m}{n}}$ there are infinitely many possible choices of $\ket{\AState{m}{n}}$, and all the states forming, e.g., the lowest energy band in \FigRef{fig:DispersionRelation} can be constructed from the \emph{same} state in a quite tautological manner, by first adding all these states together in a superposition with arbitrary non-vanishing expansion coefficients, and subsequently applying the operators (\ref{eq:TauknDefinition}) to extract all the eigenstates one by one.
This is merely a formal property of states obeying a symmetry, and hence it does not provide a physical explanation of the spatial correlations in the states.

Rather than forming the underlying state $\ket{\AState{m}{n}}$, from the energy eigenstates in each excitation band, we will build their formation on the physical principle, that they are well-localized on the lattice and centered at the site $j = 0$. Since each of the equivalence classes introduced in (\ref{eq:DefinitionOfEquivalenceClasses}) contains states which are related to each other by displacement, they all project onto the same state under the operation by
$\Tauk[n]$, which in turn implies, that we can build any quasi-momentum eigenstate $\ket{\EState{m}{n}}$ from~(\ref{eq:TranslationSuperposition}) with $\ket{\AState{m}{n}}$ composed of only one member from each equivalence class
\begin{align*}
  \ket{\EqBasis{1}{a_1}} &\in \Equivalence^{(1)}
  , &
  \ket{\EqBasis{2}{a_2}} &\in \Equivalence^{(2)}
  , &
  & \ldots
  , &
  \ket{\EqBasis{P}{a_P}} &\in \Equivalence^{(P)}.
\end{align*}

\begin{figure*}[t]
  \centering
  \includegraphics[width = \textwidth]{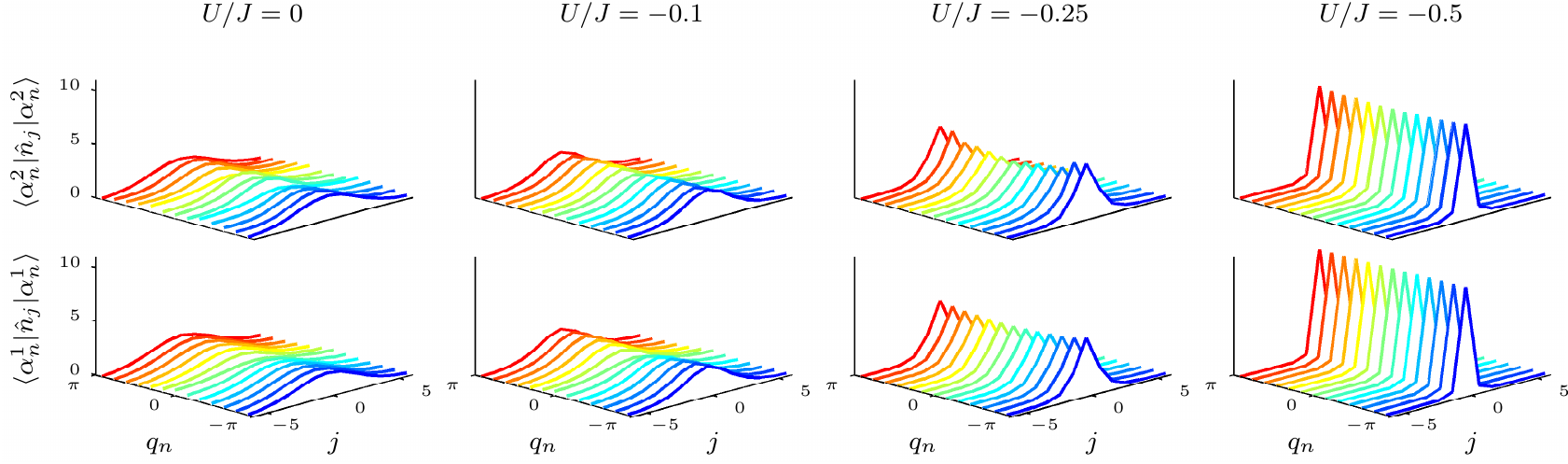}
  \caption{(Color online) Densities of $\ket{\AState{m}{n}}$ for the two lowest energy excitations ($m = 1,2$) for four different attractive interaction strengths. The rightmost panels correspond to the two lowest excitations in \FigRef{fig:DispersionRelation}(c).}
  \label{fig:AlphaDensities}
\end{figure*}

We now select the representatives of each equivalence class according to their spatial location on the lattice. Due to the periodicity of the system we can picture the sites lying in a circle as in \FigRef{fig:SchematicConcept}, and thus assign a two-dimensional position $(x_j,y_j)$ to each of the $M$ sites. For convenience we use complex notation $z_j=x_j+ \imag y_j = \Exp{2 \pi \imag j /M}$. We can now define a center-of-mass coordinate as the arithmetic mean of the coordinates in the complex plane.
This definition of the center of mass has the advantage, that when we apply the translation operator on a basis state, we rotate the complex phase angle of all particles by $2 \pi/M$ radians counter-clockwise---and the center of mass thus rotates by the same angle in the complex plane. This rotation, couples the different elements within each equivalence class (\ref{eq:StatesOfAnEquivalenceClass}), and we propose to expand our candidate localized state on members from each equivalence class which have their center-of-mass complex phase angle within the interval $] -\pi/M ; \pi/M]$. \FigRef{fig:Wedge} depicts the center-of-mass coordinate for all Fock states with 11 sites and 4 particles, and the shaded wedge indicates the values for which we include the state in our expansion.
In practice for the state vectors $\ket{\EState{m}{n}}$ found by our numerical diagonalization, we merely retain the expansion coefficients on the basis states with center-of-mass in the wedge. To summarize, we choose the underlying state as
\begin{align}
  \ket{\AState{m}{n}}
  &= \frac{\WedgeProj \ket{\EState{m}{n}}}{\sqrt{\braket{\EState{m}{n} | \WedgeProj | \EState{m}{n}}}}
\end{align}
where $\WedgeProj$ is the projection onto the space spanned by the Fock states with center-of-mass \emph{within} the wedge, and even superposition states of those Fock state pairs with center-of-mass on the upper and lower angular limits of the wedge.

\begin{figure}[b]
  \centering
  \includegraphics[width=\columnwidth]{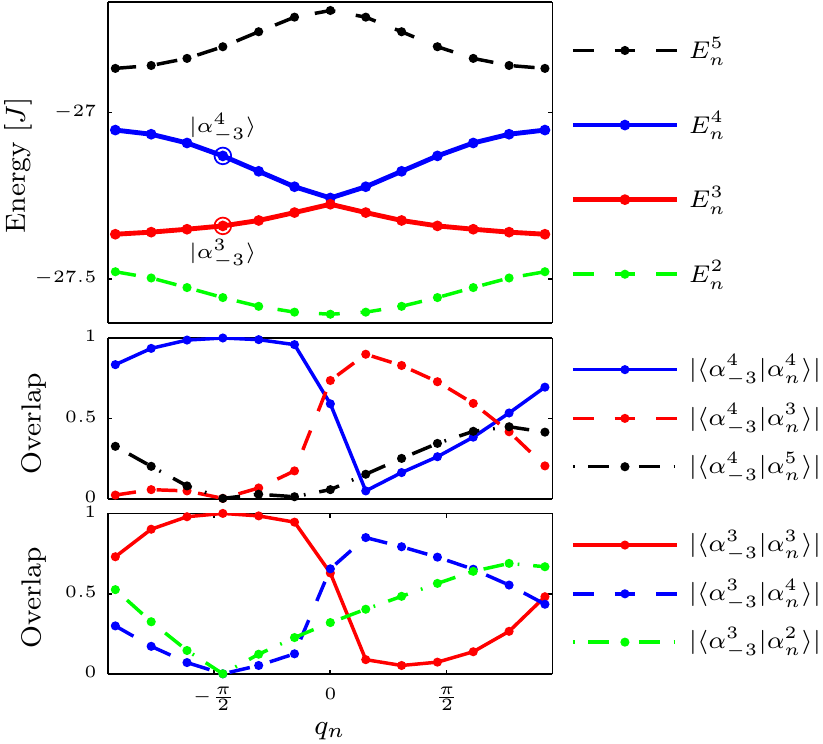}
  \caption{(Color online) Energy levels and behavior of localized states near avoided crossings. The upper panel shows the energy levels of the 2nd to 5th excitations at $U/J = - 0.45$. In the middle and lower panels we show the overlap of the localized states identified for each quasi-momentum with the one for the definite value $n_0 = -3$ in the 4th and 3rd excitations, respectively, marked with circles in the upper panel. }
  \label{fig:AlphaOverlap}
\end{figure}

\section{Localized states}
\label{sec:LocalizedStates}
In the previous section we motivated a procedure to find localized states, $\ket{\AState{m}{n}}$, which yields the energy eigenstate, $\ket{\EState{m}{n}}$, by the translation and superposition Ansatz~(\ref{eq:TranslationSuperposition})
\begin{align}
  \ket{\EState{m}{n}} = \frac{\Tauk[n]\ket{\AState{m}{n}}}{\sqrt{\braket{\AState{m}{n} |\Tauk[n] | \AState{m}{n}}}} .
\end{align}%
\FigRef{fig:AlphaDensities} shows the densities of these for four different interaction strengths and for the two lowest energy excitation numbers, $m = 1,2$. For low interaction strength, tunneling is significant, which causes the states to spread out, while for larger attraction the states narrow to minimize energy, and in the limit $U/J \to -\infty$ all particles are located on one site in the $m=1$ states. The figure also shows, that even though the states $\ket{\AState{m}{n}}$ are determined independently for each state $\ket{\EState{m}{n}}$, they are very similar for different $q_n$, supporting our assumption that the center-of-mass and the internal relative motion of the atoms separate.

In our previous discussion we justified the separation of different motional degrees of freedom by their different energy scales. Parts (a) and (b) of \FigRef{fig:DispersionRelation} lend no support for the separation, while parts (c) and (d) display a clear distinction between the energy scales associated with the center-of-mass motion and the relative motion. A magnified view of excitations 2-5 in \FigRef{fig:DispersionRelation}(c) are reproduced in the upper panel of \FigRef{fig:AlphaOverlap} (for a slightly different interaction strength). Among these states an avoided crossing is observed between the 3rd and 4th excitation bands at zero quasi-momentum. This suggests that the separation Ansatz does not hold here, and indeed, the localized states identified for these states vary, as we pass the crossing region.

This is quantified in the two lower panels of \FigRef{fig:AlphaOverlap}, where the overlap of the localized states $\ket{\AState{m}{n}}$ obtained from different quasi-momentum states with a fixed localized state $\ket{\AState{m}{-3}}$ is plotted. The solid blue curve in the middle panel shows that the localized states are almost constant for negative quasi-momentum states in the 4th excitation band with a near unit overlap with the typical state $\ket{\AState{4}{-3}}$ in that interval. However, the states abruptly changes character at the crossing, and acquires the character of the negative quasi-momentum states in the 3rd excitation band as seen from the dashed blue curve in the lower panel. Consistently, the same behavior is observed for the states in the 3rd excitation band. This is a typical example of the diabatic transition between pairs of states near degeneracies in a physical system, and we may in fact imagine an experimental application of this crossing phenomenon, where the system is driven through the avoided crossing by acceleration of the lattice, and thus superpositions of two different many-body quantum states may be controllably prepared.

The curves in the upper panel suggests that further avoided crossing with the 2nd and 5th excitation bands occur at the band edges---but more pronounced in the spectra for weaker attraction---and the overlap amplitudes in the lower panels confirm a non-adiabatic transition behavior toward these states.

\subsection{Many-body character of the localized states}

We have shown, that the energy eigenstates in the Bose-Hubbard model in the case of strong attractive interactions can be written as translational superposition states of localized composite many-body states of the atoms. 
 Let us now investigate their many-body characters in more detail.

\begin{figure}[b]
  \centering
  \includegraphics[width = \columnwidth]{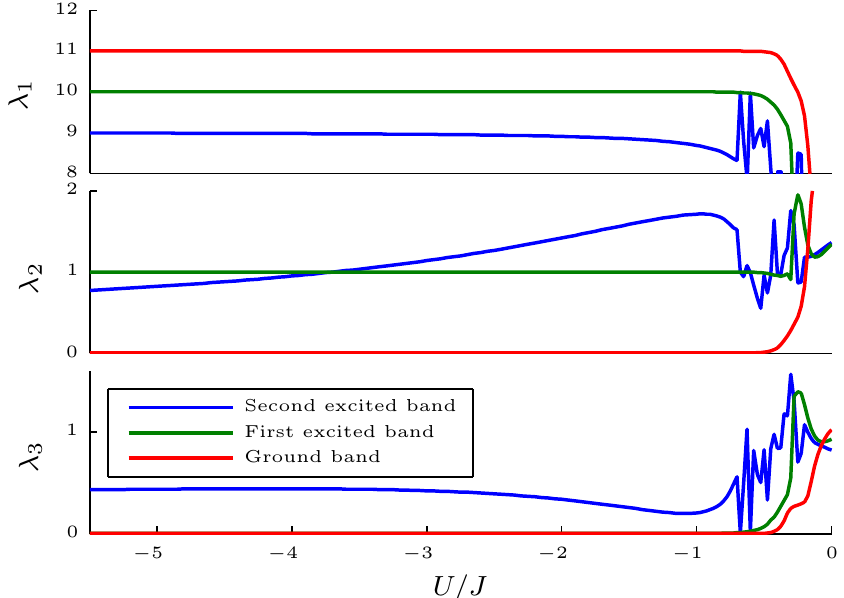}
  \caption{(Color online) The three largest eigenvalues ($\lambda_1 \geq \lambda_2 \geq \lambda_3$) of the one-body density matrix~(\ref{eq:OBDM}) for the localized states $\ket{\AState{m}{0}}$ with the lowest energy in the ground band, the first and the second energy band. The limiting value of $\lambda_1$ for strong attraction equals the number of particles, $N = 11$, in  for the ground band. The rich structures in the eigenvalues at weak interactions are due multi-level intraband level crossings which change the character of the lowest eigenstates in the first and second energy bands. }
  \label{fig:OBDMEigenvalues}
\end{figure}

For each state $\ket{\AState{m}{n}}$ we can compute the one-body density matrix $\OBDM$ with the matrix elements
\begin{align}
  \OBDM_{jk} &\defi \braket{\AState{m}{n} | \aOp_j \aNed_k | \AState{m}{n}} .
  \label{eq:OBDM}
\end{align}
This quantity is equivalent to the first order coherence function, but unlike the results shown in \FigRef{fig:CorrelationFunctions}, here we compute it for the localized state. In \FigRef{fig:OBDMEigenvalues} is plotted the three largest  eigenvalues of $\OBDM$ for the states $\ket{\AState{m}{n}}$ with $n=0$ and $m$ corresponding to the ground band and to the lowest sub-band of the first and second excited band. We we see that the largest eigenvalue of $\OBDM$ is comparable to the number of particles, which implies that in each state $\ket{\AState{m}{n}}$ many of the atoms occupy the same single particle wave function, which in turn is given by the corresponding eigenvector of $\OBDM$.
In the limit of strong attraction $U/J \to - \infty$, for the ground state the largest eigenvalue $\lambda_1$ approaches the number of particles $N=11$, while for the states in the first and second excited bands, the largest eigenvalue approaches $N-1$ and $N-2$, respectively. This is a clear indication, that in the ground band, $\ket{\AState{1}{n}}$ is an $N$-particle condensate with all particles in some single particle state $\ket{\GroundState{1}{n}}$, whose density is easily extracted from \figurename~\ref{fig:AlphaDensities}, while in the first and second excited band, one and two atoms are excited out of the condensate, respectively. This is similar to the understanding offered by the (number-conserving) Bogoliubov treatment of interacting bosons \cite{Castin} in the limit of particle-like excitations out of the condensate wave function.

\begin{figure}[t]
  \centering
  \includegraphics[width = \columnwidth]{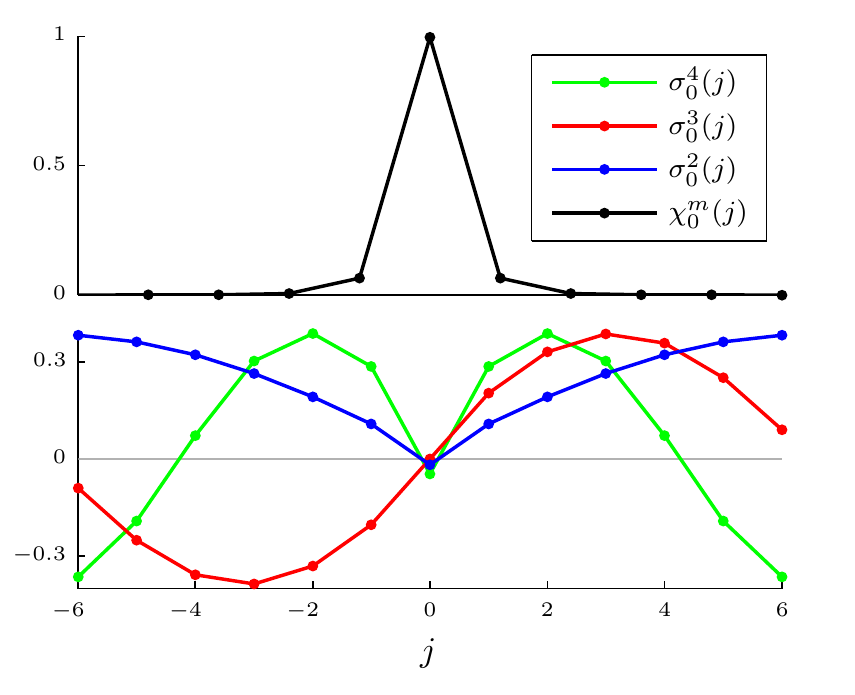}
  \caption{Wave functions of the single-particle states found as eigenvectors of the one-body density matrix~(\ref{eq:OBDM}) for $\ket{\AState{m}{n}}$ with $n = 0$ and different values of $m$ and with $U = -1.75$. Similar results are found for all values of $n$, but not with real-valued wave functions.
The upper panel shows the condensate wave-function which are practically indistinguishable for different values of $m$ in the first band, and in the lower panel we see the wave function of the single excited particle in the three lowest states in the first energy band. }
  \label{fig:FirstExcitedBandSignelParticleStates}
\end{figure}

The middle panel of \FigRef{fig:OBDMEigenvalues} shows that in the first excited band, the second largest eigenvalue of $\OBDM$ is unity for a wide range of strong attraction, suggesting that a single excited atom occupies a definite single particle state $\ket{\FirstExcited{m}{n}}$  orthogonal to the condensate mode $\ket{\GroundState{m}{n}}$. Qualitatively, this particle tunnels among $M$ lattice sites, exposed to an attractive potential dip at the site occupied by the condensate, and we expect the particle to have discrete eigenstates equivalent to plane waves with a suppressed density at the condensate location. The condensate wave function and this excited state wave function are indeed available as the eigenvectors of $\OBDM$ corresponding to the eigenvalues $\lambda_1$ and $\lambda_2$, respectively, and in \FigRef{fig:FirstExcitedBandSignelParticleStates}, we show these functions for $U/J = -1.75$ for the three lowest excitations in the first excited energy band ($m = 2,3,4$). The condensate wave function shown in the upper panel is strongly peaked with small wings due to the tunneling, and the singly populated wave functions for the three lowest states of the first excited band are indeed of the expected form with an increasing number of nodes corresponding to scattering states on the ring in the presence of the condensate.

\begin{figure}[t]
  \centering
  \includegraphics[width = \columnwidth]{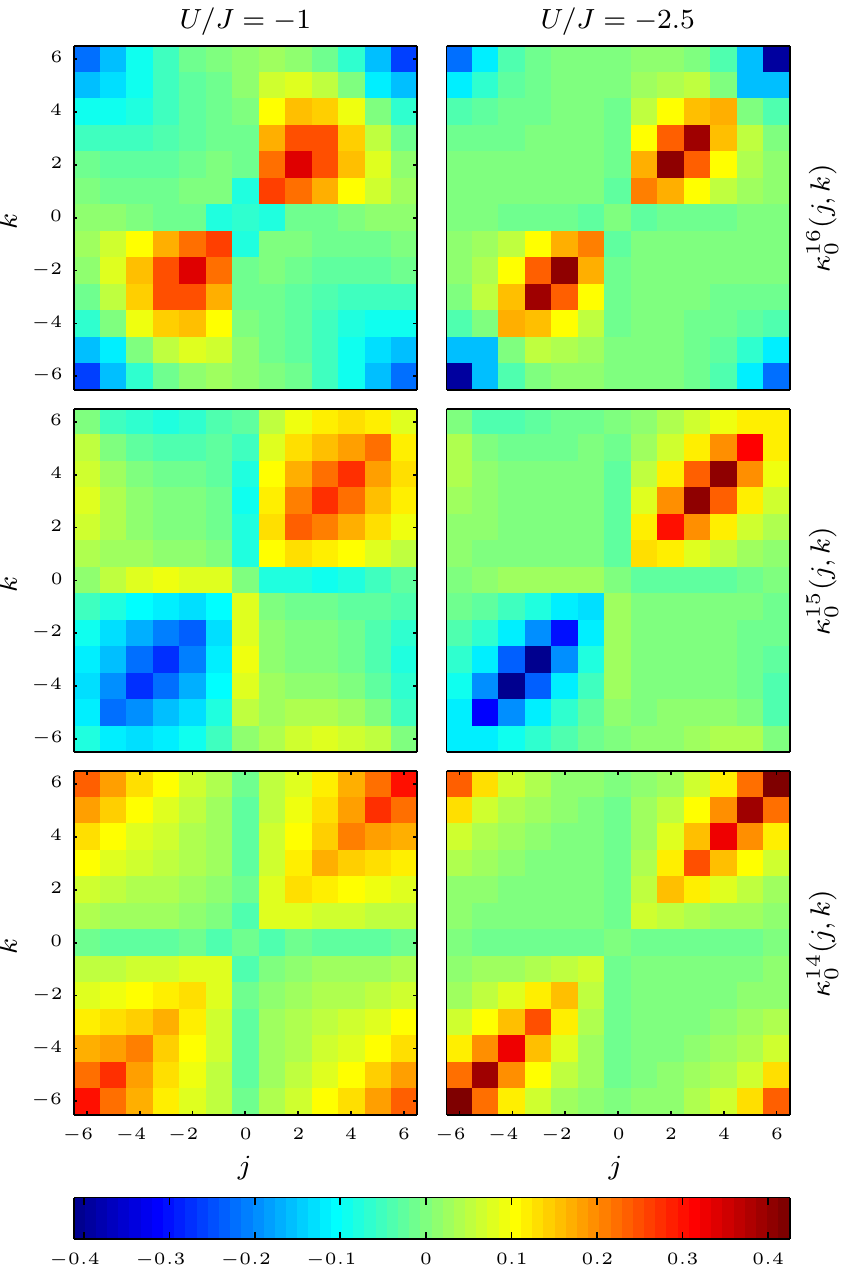}
  \caption{(Color online) Wave functions of the two particles excited out of the localized condensate state. For $n = 0$ the wave function is real-valued and the two-particle wave function amplitude is indicated by the color coding.}
  \label{fig:SecondExcitedBandTwoParticleStates}
\end{figure}

For states in the second band the largest eigenvalue of $\OBDM$ has the limiting value $N-2$, but the two subsequent eigenvalues show a more complicated behavior. This is compatible with a condensate state with $N-2$ atoms, while the two excited atoms populate a non-trivial two-particle state. To investigate this state further, 
$\Op b_c$ is applied $N-2$ times to $\ket{\AState{m}{n}}$, where $\Op b_c$ is the annihilation operator that removes a particle in the condensate mode $\ket{\GroundState{m}{n}}$.
The wave function of the resulting two-particle entangled state $\ket{\SecondExcited{m}{n}}$ is illustrated in \FigRef{fig:SecondExcitedBandTwoParticleStates} for the three lowest zero quasi-momentum states in the second excited manifold. The states are shown for two different interaction strengths and are evidently not (symmetrized) product states and can therefore not be regarded  as two particles occupying two (different) single-particle states. Indeed, the attractive interaction strongly correlates the atoms on the same side of the condensate peak at $j=0$, and as the attraction increases the wave-function becomes more and more narrow around the diagonal in the plot, indicating an increased spatial bunching of the two particles.

We note that the conventional Bogoliubov treatment is formally an expansion in $1/\sqrt{N}$  \cite{Castin}, and  interactions between atoms outside the condensate wavefunctions can be neglected in the limit of large $N$. Due to the near degeneracy of non-condensate single particle states shown in \FigRef{fig:FirstExcitedBandSignelParticleStates}, these interactions are significant in our problem and cause the emergence of the non-integer occupancy eigenvalues shown in \FigRef{fig:OBDMEigenvalues}, even when precisely two atoms are removed from the condensate state.

\section{Energy of translational motion}
\label{sec:EnergyOfTranslationalMotion}
In the previous sections we justified the separation of the quantum state into a center-of-mass translational motion and relative motion of the atoms. We observed how the discrete excitations of the system in the case of strong attraction correspond to removal of one or more atoms from the condensate-like bound state of the system, while the lowest energy scale is associated with translation of the compound state.

The localized state breaks the translational symmetry of the problem, while the restoration of the symmetry by a superposition of translated states gives rise to a simple analysis and interpretation of the ensuing energy. In continuous space, Peierls and Yoccoz \cite{Peierls}, thus show that a superposition state with a complex phase variation with wave number $K$ is a total momentum eigenstate and gives rise to the usual kinetic energy expression, $\hslash^2K^2/2M$, where $M$ is the total mass of the system of particles. Peierls and Yoccoz also show that for an oriented quantum system, an angular superposition state weighted by Wigner rotation functions becomes an angular momentum eigenstate, e.g. of a nucleus or a molecule. In this case, however, the spectrum only recovers the rigid rotor form with terms proportional to $L(L+1)$ and $M_L^2$ \cite{Herzberg} if the state is sufficiently localized in Euler angles such that a Taylor expansion of the Wigner rotation functions is valid over the angular extent of the state. In nuclei, a superfluid core violates this, but the normal component may still be treated in the above sense and gives rise to rotational structure with an apparently reduced moment of inertia.

In case of angular rotation of polarized electronic states in atoms, deviations from the rigid rotor energy structure in the form of an apparent \textit{increased} moment of inertia, have been explained in \cite{Larsrot}, by the angular extent of the states being too wide for the simple Taylor expansion of the Wigner function to be valid.

The case of a finite, discrete lattice presents an interesting intermediate case between the continuous momentum and discrete angular momentum eigenstates, and the dispersion relation for center-of-mass motion in the lattice is not a priori obvious.

By our numerical diagonalization, we have already found the spectrum, but following \cite{Peierls}, we can also evaluate the energy dispersion with direct reference to the representation of the translation and energy eigenstates $\ket{\EState{m}{n}}$ in terms of the underlying state $\ket{\AState{m}{n}}$ given by (\ref{eq:TranslationSuperposition}). This leads to the expectation value of $\Op H$,
\begin{align}
  E^m_n
  &= \braket{\EState{m}{n} | \Op H | \EState{m}{n}}
  = \frac{\braket{\AState{m}{n} | \Op H \Tauk[n] | \AState{m}{n}}}{\braket{ \AState{m}{n} | \Tauk[n] | \AState{m}{n} }} ,
  \label{eq:EnergyInTermsOfAlpha}
\end{align}
where we have used that $\Tauk[n]$ commutes with $\Op H$. Inserting the definition (\ref{eq:TauknDefinition}) into the numerator yields
\begin{align}
  \braket{ \AState{m}{n} | \Op H \Tauk[n] |  \AState{m}{n}}
  &= \frac{1}{M} \sum_{k = 0}^{M-1} \Exp{\imag k q_n} \braket{ \AState{m}{n} | \Op H \Translation^k |  \AState{m}{n}}
  \label{eq:EnergyNumerator}
\end{align}
and a similarly expression with $\Op H$ replaced by $\Identity$ applies for the normalization factor in the denominator in (\ref{eq:EnergyInTermsOfAlpha}).

Since the Hamiltonian favors nearest neighbor interactions, and since the states $\ket{\AState{m}{n}}$ become very narrow in the limit $U/J \to - \infty$, we expect that only the terms $k \in \tuborg{M-1 ,0 , 1}$ contribute significantly in the sum (\ref{eq:EnergyNumerator}). For the denominator we can make an even stronger argument for the suppression of terms: Since $\ket{\AState{m}{n}}$ is projection onto the Fock basis states with center of mass in the wedge around the positive part of the real axis in the complex plane (see \FigRef{fig:Wedge}), and since $\Translation$ makes the center of mass rotate by the angle $2\pi/M$ in the complex plane, all terms
$\braket{ \AState{m}{n} | \Translation^k |  \AState{m}{n}}$
with $k = 2 , \ldots , M-2$ vanish identically, and terms with $k =1$ or $k = M-1$ get their only contributions from Fock basis states components of $\ket{\AState{m}{n}}$ with their center of mass phase angle \emph{exactly} equal to $\pm\pi/M$.
We recall that this very strong truncation is valid because we chose to represent our states by Fock states with their center-of-mass coordinate within the wedge.

Truncating~(\ref{eq:EnergyNumerator}) to $k=M-1,0,1$ and approximating the denominator in ~(\ref{eq:EnergyInTermsOfAlpha}) by unity yields
\begin{align}
  E^m_n
  &\approx \braket{ \AState{m}{n} | \Op H |  \AState{m}{n}} + 2 \Re \kantpar{\Exp{\imag  q_n} \braket{ \AState{m}{n} | \Op H \Translation |  \AState{m}{n}}} ,
  \label{eq:EnergyApproximation}
\end{align}
which turns out to be in excellent agreement with the energy eigenvalues found from the numerical diagonalization of $\Op H$, and our description thus accounts for the spectral features at all observable energy scales.

\section{Conclusion}
\label{sec:Conclusion}
In conclusion, we have solved the attractive Bose-Hubbard Hamiltonian numerically on a ring lattice with a finite number of atoms, and we have found that while the eigenstates have uniform mean population on all sites, the atomic wave function shows strong interparticle correlations. These correlations are suggestive that the exact eigenstates can be represented as superposition states of translated replica of compact (solitonic) composite many-body states. We have verified that this representation is, indeed, a useful one, and we have further demonstrated that the separation of center-of-mass and relative motion of the atoms explains features of the energy spectra very well. The argument is similar to the argument for the Born-Oppenheimer approximation applied in chemistry to separate nuclear and electronic motion. The separation is valid in the limit of strong interactions, where the energy scales become sufficiently different, and we observed how degeneracies in the system give rise to typical anti-crossing behavior with significant non-adiabatic coupling of different internal states. This phenomenon is also well known in Born-Oppenheimer theory, and plays a prominent role in the rich physics associated with conical intersections \cite{RevModPhys.68.985}.

Our analysis supplements numerical calculations with a physical picture of the states of attractive bosons on a lattice. In this picture they form compact, macroscopically populated condensate states, and they may be subject to excitations of one or several atoms out of the condensate state, which in turn occupies a superposition of different locations corresponding to center-of-mass quasi-momentum eigenstates of the system.

Apart from offering insight in the anatomy of the many-body quantum state, the separation of the degrees of freedom in relative and center-of-mass motion paves the way for new mechanisms to excite and control the system dynamics. We imagine, for example, that the non-adiabatic coupling of different many-body states across a certain range of quasi-momentum eigenstates, may be applied to prepare interesting many-body superposition states.

\appendix

\section{Ordering of Fock basis states}
\label{sec:Ordering}

In order to diagonalize a Hamiltonian which is given in a second quantization formulation as
\begin{align} \label{eq:A1}
  \Op H &= \sum_{s,t} c_{st} \aOp_s \aNed_t + \sum_{s,t,u,v} d_{stuv} \aOp_s \aOp_t \aNed_u \aNed_v
\end{align}
where the summation indices run over all the single particle modes, we must write it as a matrix in a suitable basis, e.g., multi-mode Fock states, of many-body states:
\begin{align*}
  H_{ab} &\defi \braket{\Obasis{a} | \Op H | \Obasis{b}} .
\end{align*}
Instead of iterating through all pairs of basis states, a lot can be saved by using the sparsity of $\Op H$, which only couples one basis state $\ket{\Obasis{b}}$ to a few others.

For single particle operators we thus apply $\aOp_s \aNed_t$ to $\ket{\Obasis{b}}$ (for all $s$ and $t$ with non-vanishing $c_{st}$ in (\ref{eq:A1})), and the resulting state can then be identified among the basis states. A similar procedure is used for the two-particle operators, but here one iterates over four indices instead of two. The identification procedure can be done conveniently with a suitable hashing function that facilitates the handling of single mode indices in the full set of many-body states. Different general hashing functions are suggested in \cite{ZangDong}, but in the case where the Fock basis consists of all states with $N$ bosons in $M$ modes, \cite{ShoudanLiang} proposes a simpler hashing function that gives a convenient mapping between the single mode occupation numbers in the Fock basis states and the set of integers from $1$ to $\TotNum{M}{N}$. In \cite{ShoudanLiang} only the closed form formula for the hashing function is given, but here we define a binary relation to compare each two basis states and we derive the hashing formula from first principles. Finally, we show that, using the ordering induced by this particular binary relation, it is often not even necessary to use the hashing formula.

\subsection{Binary relation}

For two multi-mode $N$-particle Fock states
\begin{align*}
  \ket{\Obasis{a}} &= \ket{m_0 , \ldots , m_{M-1}}
  \\
  \ket{\Obasis{b}} &= \ket{n_0 , \ldots , n_{M-1}}
\end{align*}
there will be a lowest $k \in \tuborg{0 , \ldots , M-1}$ such that
\begin{align*}
  m_j = n_j \quad \text{for} \quad j > k .
\end{align*}
We now introduce a binary relation $\leqN{N}$ on the basis,
\begin{gather}
  \begin{aligned}
    n_k \leq m_k
    \quad &\iff \quad
    \ket{\Obasis{b}} \leqN{N} \ket{\Obasis{a}} ,
    \\
     n_k \geq m_k
    \quad &\iff \quad
    \ket{\Obasis{a}} \leqN{N} \ket{\Obasis{b}} .
  \end{aligned}
  \label{eq:BinaryRelation}
\end{gather}
This relation provides an ordering of the full set of basis states, and a simple recipe to find the successor of a given basis state, $\ket{\Obasis{a}} = \ket{n_0 , \ldots , n_{M-1}}$:
First find the highest $k \in \tuborg{0 , \ldots , M-1}$ such that
\begin{align*}
  n_j = 0 \quad \text{for} \quad j < k ,
\end{align*}
i.e. the state has the form
\begin{align*}
  \ket{\Obasis{a}} &= \ket{0 , \ldots , 0 , n_k , n_{k+1} , \ldots , n_{M-1}} .
\end{align*}
with $n_k \neq 0$. The next element in the ordered basis is then obtained by transferring one particle from the $k$th to the $k+1$st mode, while the remaining particles in the $k$th mode are moved to the first one:
\begin{align*}
  \ket{\Obasis{a+1}} &= \ket{n_k - 1 , \ldots , 0 , 0 , n_{k+1} + 1 , \ldots , n_{M-1}} .
\end{align*}
This increment recipe clearly respects the ordering, $\ket{\Obasis{a}} \leqN{N} \ket{\Obasis{a+1}}$, and it can be proved inductively that successive use of this increment constructs all possible basis states. Thus, from now on we let
\begin{align*}
  \FockBasisStates_N^M = \af{\ket{\Obasis[N]{1}} , \ket{\Obasis[N]{2}} , \ldots , \ket{\Obasis[N]{\TotNum{M}{N}}}}
\end{align*}
denote the \emph{ordered} set of $N$-particle Fock basis states ($N = 0 , 1 , 2 , \ldots$) with the ordering induced by (\ref{eq:BinaryRelation}).

\subsection{Explicit enumeration of basis states}

In our calculations we make explicit use of the translational invariance of the problem, and hence we also need efficient means to identify the Fock state members of the equivalence classes $\tuborg{\Equivalence^{(j)}}_{j = 1}^P$, and the action of the translation operator $\Translation$. The action of $\Translation$ on a multi-mode Fock state is trivially carried out by shifting the occupation numbers cyclically, but the problem lies in identifying which basis state one arrives at. For this purpose it is advantageous to be able to calculate the basis state number directly (in the chosen ordering) without searching in a complete list of basis states.

For this purpose, it is advantageous to represent the states by the $N$-tuple $(s_1 , \ldots , s_N)$, where $s_j \in \tuborg{0 , \ldots , M-1}$ denotes which mode is occupied by the $j$th particle. This is a unique representation and it is not in conflict with the particle indistinguishability, if we choose $s_1 \leq s_2 \leq \cdots \leq s_N$.

To each  $N$-tuple corresponds a basis state with a given number $\Order{N}(s_1 , \ldots , s_N) \in \tuborg{1,\ldots , \TotNum{M}{N}}$ in our chosen ordering, and to identify that number efficiently we note that
%
%
the ordering begins with the states where we occupy only the first mode, then comes all possible ways of occupying only the first two modes, but with non-vanishing occupation of the second mode, then the first three modes and so forth. The last state which only occupies the first $s$ modes $0 , \ldots , s-1$ is the state
\begin{align}
  \ket{0 , \ldots , 0  , N , 0 , \ldots , 0}
\end{align}
where the $N$ appears in the place corresponding to mode $s-1$, and this state must have the number $\TotNum{s}{N}$. The subsequent states, according to our ordering, start by putting one particle in the next mode and the remaining $N-1$ atoms in lower modes, and these states, all having one atom in the mode corresponding to the $s$th mode, are to be numbered as $\TotNum{s}{N}$ plus the sequential enumeration of the $N-1$ particles into the modes $0, \ldots , s-1$:
\begin{align*}
  \Order{N}(s_1 , \ldots , s_{N-1} , s) = \TotNum{s}{N} + \Order{N-1}(s_1 , \ldots , s_{N-1}) ,
\end{align*}
where $s_1 \leq \ldots \leq s_{N-1} \leq s$. By inserting this recursive relation into itself repeatedly we end up at the last term
\begin{align*}
  \Order{1}(s_1) = s_1 + 1 = \TotNum{s_1}{1} + 1
\end{align*}
(where we have extended the binomial coefficient such that $\binom{n}{k} = 0$ when $k > n \geq 0$). This proves the explicit number formula for the hashing function
\begin{align}
  \Order{N}(s_1 , \ldots , s_N)
  &= 1 + \sum_{j = 1}^N \TotNum{s_j}{j} .
  \label{eq:HashingFunction}
\end{align}
Formula (\ref{eq:HashingFunction}) equips us with an explicit way of calculating the number of the basis vector from its $N$-tuple representation, and we can hence readily calculate the action of, e.g., the translation operator without having to search through the ordered list of states for a match of the shifted populations.

\subsection{Matrix elements}
The great advantage of using this ordering of the basis states is that, even though we just derived the explicit formula (\ref{eq:HashingFunction}), often we need not even calculate the hashing function. 


If we apply an annihilation operator $\aNed_j$ to an $N$-particle Fock basis state $\ket{\Obasis[N]{a}}$, with a non-vanishing occupation in the $j$th mode, $\braket{\Obasis[N]{a} | \Op n_j | \Obasis[N]{a}} \neq 0$, the resulting state  $\aNed_j \ket{\Obasis[N]{a}}$ is proportional to an $N-1$-particle Fock basis state, while $\aNed_j \ket{\Obasis[N]{a}}$ vanishes  if $\braket{\Obasis[N]{a} | \Op n_j | \Obasis[N]{a}} = 0$. But the ordering has the property that if two $N$-particle states that are related as $\ket{\Obasis[N]{a}} \leqN{N} \ket{\Obasis[N]{b}}$ both occupy the $j$th mode, then the two $N-1$-particle states obtained by operating with $\aNed_j$ have the same ordering:
\begin{align}
  \frac{\aNed_j\ket{\Obasis[N]{a}}}{\sqrt{\braket{\Obasis[N]{a} |\Op n_j | \Obasis[N]{a}}}} 
  \leqN{N-1}
  \frac{\aNed_j\ket{\Obasis[N]{b}}}{\sqrt{\braket{\Obasis[N]{b} |\Op n_j | \Obasis[N]{b}}}}  .
  \label{eq:OrderPreserved}
\end{align}
There is, indeed, a bijective correspondence between the ordered subset of basis states with non-vanishing occupation of the $j$th mode ($j = 0 , \ldots , M-1$)
\begin{align*}
  \Occupation^N_j &= \af{\ket{\Obasis[N]{p_1^{(j)}}} , \ldots , \ket{\Obasis[N]{p_s^{(j)}}} , \ldots , \ket{\Obasis[N]{p_{\TotNum{M}{N-1}}^{(j)}} }}  \subset \FockBasisStates_N^M
\intertext{and the $N-1$ particle multi-mode basis}
\FockBasisStates_{N-1}^M &= \af{\ket{\Obasis[N-1]{1}} , \ldots , \ket{\Obasis[N-1]{s}} , \ldots , \ket{\Obasis[N-1]{\TotNum{M}{N-1}} }}
\end{align*}
and since $\aNed_j$ (restricted to $\Occupation^N_j$) conserves the ordering of the Fock basis states in the sense~(\ref{eq:OrderPreserved}), we conclude that for all $\ket{\Obasis[N]{p_s^{(j)}}} \in \Occupation^N_j$ we have
\begin{align*}
  \aNed_j\ket{\Obasis[N]{p_s^{(j)}}} &= \sqrt{\braket{\Obasis[N]{p_s^{(j)}} | \Op n_j | \Obasis[N]{p_s^{(j)}}}} \times  \ket{\Obasis[N-1]{s}} .
\end{align*}
Therefore, if we take two Fock basis states with occupation of the $j$th and $k$th mode, respectively, $\ket{\Obasis[N]{p_s^{(j)}}} \in \Occupation^N_j$ and $\ket{\Obasis[N]{p_t^{(k)}}} \in \Occupation^N_k$, we get
\begin{align*}
  \braket{\Obasis[N]{p_s^{(j)}} | \aOp_j \aNed_k | \Obasis[N]{p_t^{(k)}}}
  &= C_{st}^{jk} \braket{\Obasis[N-1]{s} | \Obasis[N-1]{t}}
  = C_{st}^{jk} \delta_{s,t}
\end{align*}
where the factor of proportionality is found from the occupation numbers of the two states
\begin{align}
  C_{st}^{jk} &= \sqrt{\braket{\Obasis[N]{p_s^{(j)}} | \Op n_j | \Obasis[N]{p_s^{(j)}}}
    \times \braket{\Obasis[N]{p_t^{(k)}} | \Op n_k | \Obasis[N]{p_t^{(k)}}} } .
  \label{eq:NumericalFactor}
\end{align}
%
%

The matrix representation of the operator $\aOp_j \aNed_k$ is thus found by identifying \emph{all} the Fock basis states that have non-zero occupation of the $j$th and $k$th mode, respectively, while keeping their individual order. These states are indexed by the lists of numbers,  $\{p^{(j)}_s\}_{s=1}^{\TotNum{M}{N-1}}$ and $\{p^{(k)}_t\}_{t=1}^{\TotNum{M}{N-1}}$, and then the only non-vanishing entries in the matrix representation of $\aOp_j \aNed_k$ are indexed as $(p^{(j)}_s , p^{(k)}_s)$ for $s = 1,\ldots ,\TotNum{M}{N-1}$, where we must calculate the numerical factors (\ref{eq:NumericalFactor}) from the occupations of the two modes in question for all the pairs of basis states.

The first term in the Hamiltonian (\ref{eq:A1}) is merely a linear combination of such operators, so the matrix representation of that term follows easily from the above. The interaction terms with application of two annihilation operators and two creation operators benefit from the same use of ordered sets and convenient enumeration of states with non-vanishing occupancies. In the preparation of the Hamiltonian matrix it is hence not even necessary to evaluate or utilize the full hashing function following from the preceding subsection.

\bibliography{BHreferencer}

\begin{thebibliography}{29}
\expandafter\ifx\csname natexlab\endcsname\relax\def\natexlab#1{#1}\fi
\expandafter\ifx\csname bibnamefont\endcsname\relax
  \def\bibnamefont#1{#1}\fi
\expandafter\ifx\csname bibfnamefont\endcsname\relax
  \def\bibfnamefont#1{#1}\fi
\expandafter\ifx\csname citenamefont\endcsname\relax
  \def\citenamefont#1{#1}\fi
\expandafter\ifx\csname url\endcsname\relax
  \def\url#1{\texttt{#1}}\fi
\expandafter\ifx\csname urlprefix\endcsname\relax\def\urlprefix{URL }\fi
\providecommand{\bibinfo}[2]{#2}
\providecommand{\eprint}[2][]{\url{#2}}

\bibitem[{\citenamefont{Herzberg}(1966)}]{Herzberg}
\bibinfo{author}{\bibfnamefont{G.}~\bibnamefont{Herzberg}},
  \emph{\bibinfo{title}{Molecular spectra and molecular structure}},
  no.~\bibinfo{number}{2} in \bibinfo{series}{Molecular spectra and molecular
  structure} (\bibinfo{publisher}{Van Nostrand Reinhold},
  \bibinfo{year}{1966}).

\bibitem[{\citenamefont{Bohr and Mottelson}(1969)}]{Bohr}
\bibinfo{author}{\bibfnamefont{A.}~\bibnamefont{Bohr}} \bibnamefont{and}
  \bibinfo{author}{\bibfnamefont{B.}~\bibnamefont{Mottelson}},
  \emph{\bibinfo{title}{Nuclear structure}}, no.~\bibinfo{number}{2} in
  \bibinfo{series}{Nuclear Structure} (\bibinfo{publisher}{W. A. Benjamin},
  \bibinfo{year}{1969}).

\bibitem[{\citenamefont{M{\o}lmer and Taulbjerg}(1988)}]{Taulbjerg}
\bibinfo{author}{\bibfnamefont{K.}~\bibnamefont{M{\o}lmer}} \bibnamefont{and}
  \bibinfo{author}{\bibfnamefont{K.}~\bibnamefont{Taulbjerg}},
  \bibinfo{journal}{Journal of Physics B: Atomic, Molecular and Optical
  Physics} \textbf{\bibinfo{volume}{21}}, \bibinfo{pages}{1739}
  (\bibinfo{year}{1988}).

\bibitem[{\citenamefont{Tanner et~al.}(2000)\citenamefont{Tanner, Richter, and
  Rost}}]{Rost}
\bibinfo{author}{\bibfnamefont{G.}~\bibnamefont{Tanner}},
  \bibinfo{author}{\bibfnamefont{K.}~\bibnamefont{Richter}}, \bibnamefont{and}
  \bibinfo{author}{\bibfnamefont{J.-M.} \bibnamefont{Rost}},
  \bibinfo{journal}{Rev. Mod. Phys.} \textbf{\bibinfo{volume}{72}},
  \bibinfo{pages}{497} (\bibinfo{year}{2000}).

\bibitem[{\citenamefont{Madsen and M\o{}lmer}(2001{\natexlab{a}})}]{lithium}
\bibinfo{author}{\bibfnamefont{L.~B.} \bibnamefont{Madsen}} \bibnamefont{and}
  \bibinfo{author}{\bibfnamefont{K.}~\bibnamefont{M\o{}lmer}},
  \bibinfo{journal}{Phys. Rev. Lett.} \textbf{\bibinfo{volume}{87}},
  \bibinfo{pages}{133002} (\bibinfo{year}{2001}{\natexlab{a}}).

\bibitem[{\citenamefont{Zakharov and Shabat}(1972)}]{Shabat-Inverse-Scattering}
\bibinfo{author}{\bibfnamefont{V.~E.} \bibnamefont{Zakharov}} \bibnamefont{and}
  \bibinfo{author}{\bibfnamefont{A.~B.} \bibnamefont{Shabat}},
  \bibinfo{journal}{Sov. Phys} \textbf{\bibinfo{volume}{34}},
  \bibinfo{pages}{62} (\bibinfo{year}{1972}).

\bibitem[{\citenamefont{P\'erez-Garc\'ia
  et~al.}(1998)\citenamefont{P\'erez-Garc\'ia, Michinel, and Herrero}}]{Garcia}
\bibinfo{author}{\bibfnamefont{V.~M.} \bibnamefont{P\'erez-Garc\'ia}},
  \bibinfo{author}{\bibfnamefont{H.}~\bibnamefont{Michinel}}, \bibnamefont{and}
  \bibinfo{author}{\bibfnamefont{H.}~\bibnamefont{Herrero}},
  \bibinfo{journal}{Phys. Rev. A} \textbf{\bibinfo{volume}{57}},
  \bibinfo{pages}{3837} (\bibinfo{year}{1998}).

\bibitem[{\citenamefont{Carr et~al.}(2000)\citenamefont{Carr, Leung, and
  Reinhardt}}]{Carr}
\bibinfo{author}{\bibfnamefont{L.~D.} \bibnamefont{Carr}},
  \bibinfo{author}{\bibfnamefont{M.~A.} \bibnamefont{Leung}}, \bibnamefont{and}
  \bibinfo{author}{\bibfnamefont{W.~P.} \bibnamefont{Reinhardt}},
  \bibinfo{journal}{Journal of Physics B: Atomic, Molecular and Optical
  Physics} \textbf{\bibinfo{volume}{33}}, \bibinfo{pages}{3983}
  (\bibinfo{year}{2000}).

\bibitem[{\citenamefont{Pethick and Pitaevskii}(2000)}]{PethickPitaevskii}
\bibinfo{author}{\bibfnamefont{C.~J.} \bibnamefont{Pethick}} \bibnamefont{and}
  \bibinfo{author}{\bibfnamefont{L.~P.} \bibnamefont{Pitaevskii}},
  \bibinfo{journal}{Phys. Rev. A} \textbf{\bibinfo{volume}{62}},
  \bibinfo{pages}{033609} (\bibinfo{year}{2000}).

\bibitem[{\citenamefont{Gajda}(2006)}]{Gajda}
\bibinfo{author}{\bibfnamefont{M.}~\bibnamefont{Gajda}},
  \bibinfo{journal}{Phys. Rev. A} \textbf{\bibinfo{volume}{73}},
  \bibinfo{pages}{023603} (\bibinfo{year}{2006}).

\bibitem[{\citenamefont{Zinner and Jensen}(2008)}]{Zinner}
\bibinfo{author}{\bibfnamefont{N.~T.} \bibnamefont{Zinner}} \bibnamefont{and}
  \bibinfo{author}{\bibfnamefont{A.~S.} \bibnamefont{Jensen}},
  \bibinfo{journal}{Phys. Rev. C} \textbf{\bibinfo{volume}{78}},
  \bibinfo{pages}{041306} (\bibinfo{year}{2008}).

\bibitem[{\citenamefont{Yamada et~al.}(2008)\citenamefont{Yamada, Funaki,
  Horiuchi, R\"opke, Schuck, and Tohsaki}}]{Yamada}
\bibinfo{author}{\bibfnamefont{T.}~\bibnamefont{Yamada}},
  \bibinfo{author}{\bibfnamefont{Y.}~\bibnamefont{Funaki}},
  \bibinfo{author}{\bibfnamefont{H.}~\bibnamefont{Horiuchi}},
  \bibinfo{author}{\bibfnamefont{G.}~\bibnamefont{R\"opke}},
  \bibinfo{author}{\bibfnamefont{P.}~\bibnamefont{Schuck}}, \bibnamefont{and}
  \bibinfo{author}{\bibfnamefont{A.}~\bibnamefont{Tohsaki}},
  \bibinfo{journal}{Phys. Rev. A} \textbf{\bibinfo{volume}{78}},
  \bibinfo{pages}{035603} (\bibinfo{year}{2008}).

\bibitem[{\citenamefont{Zi\ifmmode~\acute{n}\else \'{n}\fi{}
  et~al.}(2008)\citenamefont{Zi\ifmmode~\acute{n}\else \'{n}\fi{},
  Ole\ifmmode~\acute{s}\else \'{s}\fi{}, Trippenbach, and Sacha}}]{Zin}
\bibinfo{author}{\bibfnamefont{P.}~\bibnamefont{Zi\ifmmode~\acute{n}\else
  \'{n}\fi{}}},
  \bibinfo{author}{\bibfnamefont{B.}~\bibnamefont{Ole\ifmmode~\acute{s}\else
  \'{s}\fi{}}}, \bibinfo{author}{\bibfnamefont{M.}~\bibnamefont{Trippenbach}},
  \bibnamefont{and} \bibinfo{author}{\bibfnamefont{K.}~\bibnamefont{Sacha}},
  \bibinfo{journal}{Phys. Rev. A} \textbf{\bibinfo{volume}{78}},
  \bibinfo{pages}{023620} (\bibinfo{year}{2008}).

\bibitem[{\citenamefont{Kanamoto et~al.}(2006)\citenamefont{Kanamoto, Saito,
  and Ueda}}]{Kanamoto}
\bibinfo{author}{\bibfnamefont{R.}~\bibnamefont{Kanamoto}},
  \bibinfo{author}{\bibfnamefont{H.}~\bibnamefont{Saito}}, \bibnamefont{and}
  \bibinfo{author}{\bibfnamefont{M.}~\bibnamefont{Ueda}},
  \bibinfo{journal}{Phys. Rev. A} \textbf{\bibinfo{volume}{73}},
  \bibinfo{pages}{033611} (\bibinfo{year}{2006}).

\bibitem[{\citenamefont{Sacha et~al.}(2009)\citenamefont{Sacha, M\"uller,
  Delande, and Zakrzewski}}]{Sacha}
\bibinfo{author}{\bibfnamefont{K.}~\bibnamefont{Sacha}},
  \bibinfo{author}{\bibfnamefont{C.~A.} \bibnamefont{M\"uller}},
  \bibinfo{author}{\bibfnamefont{D.}~\bibnamefont{Delande}}, \bibnamefont{and}
  \bibinfo{author}{\bibfnamefont{J.}~\bibnamefont{Zakrzewski}},
  \bibinfo{journal}{Phys. Rev. Lett.} \textbf{\bibinfo{volume}{103}},
  \bibinfo{pages}{210402} (\bibinfo{year}{2009}).

\bibitem[{\citenamefont{Weiss and Castin}(2009)}]{WeissCastin}
\bibinfo{author}{\bibfnamefont{C.}~\bibnamefont{Weiss}} \bibnamefont{and}
  \bibinfo{author}{\bibfnamefont{Y.}~\bibnamefont{Castin}},
  \bibinfo{journal}{Phys. Rev. Lett.} \textbf{\bibinfo{volume}{102}},
  \bibinfo{pages}{010403} (\bibinfo{year}{2009}).

\bibitem[{\citenamefont{Streltsov et~al.}(2011)\citenamefont{Streltsov, Alon,
  and Cederbaum}}]{Streltsov}
\bibinfo{author}{\bibfnamefont{A.~I.} \bibnamefont{Streltsov}},
  \bibinfo{author}{\bibfnamefont{O.~E.} \bibnamefont{Alon}}, \bibnamefont{and}
  \bibinfo{author}{\bibfnamefont{L.~S.} \bibnamefont{Cederbaum}},
  \bibinfo{journal}{Phys. Rev. Lett.} \textbf{\bibinfo{volume}{106}},
  \bibinfo{pages}{240401} (\bibinfo{year}{2011}).

\bibitem[{\citenamefont{Winkler et~al.}(2006)\citenamefont{Winkler, Thalhammer,
  Lang, Grimm, Hecker~Denschlag, Daley, Kantian, B\"{u}chler, and
  Zoller}}]{Winkler}
\bibinfo{author}{\bibfnamefont{K.}~\bibnamefont{Winkler}},
  \bibinfo{author}{\bibfnamefont{G.}~\bibnamefont{Thalhammer}},
  \bibinfo{author}{\bibfnamefont{F.}~\bibnamefont{Lang}},
  \bibinfo{author}{\bibfnamefont{R.}~\bibnamefont{Grimm}},
  \bibinfo{author}{\bibfnamefont{J.}~\bibnamefont{Hecker~Denschlag}},
  \bibinfo{author}{\bibfnamefont{A.~J.} \bibnamefont{Daley}},
  \bibinfo{author}{\bibfnamefont{A.}~\bibnamefont{Kantian}},
  \bibinfo{author}{\bibfnamefont{H.~P.} \bibnamefont{B\"{u}chler}},
  \bibnamefont{and} \bibinfo{author}{\bibfnamefont{P.}~\bibnamefont{Zoller}},
  \bibinfo{journal}{Nature} \textbf{\bibinfo{volume}{441}},
  \bibinfo{pages}{853} (\bibinfo{year}{2006}).

\bibitem[{\citenamefont{Valiente and Petrosyan}(2008)}]{Manuel}
\bibinfo{author}{\bibfnamefont{M.}~\bibnamefont{Valiente}} \bibnamefont{and}
  \bibinfo{author}{\bibfnamefont{D.}~\bibnamefont{Petrosyan}},
  \bibinfo{journal}{Journal of Physics B: Atomic, Molecular and Optical
  Physics} \textbf{\bibinfo{volume}{41}}, \bibinfo{pages}{161002}
  (\bibinfo{year}{2008}).

\bibitem[{\citenamefont{Nygaard et~al.}(2008)\citenamefont{Nygaard, Piil, and
  M\o{}lmer}}]{Nygaard}
\bibinfo{author}{\bibfnamefont{N.}~\bibnamefont{Nygaard}},
  \bibinfo{author}{\bibfnamefont{R.}~\bibnamefont{Piil}}, \bibnamefont{and}
  \bibinfo{author}{\bibfnamefont{K.}~\bibnamefont{M\o{}lmer}},
  \bibinfo{journal}{Phys. Rev. A} \textbf{\bibinfo{volume}{78}},
  \bibinfo{pages}{023617} (\bibinfo{year}{2008}).

\bibitem[{\citenamefont{Castin and Herzog}(2001)}]{CastinHerzog}
\bibinfo{author}{\bibfnamefont{Y.}~\bibnamefont{Castin}} \bibnamefont{and}
  \bibinfo{author}{\bibfnamefont{C.}~\bibnamefont{Herzog}},
  \bibinfo{journal}{Comptes Rendus de l'Acad{\'e}mie des Sciences --- Series IV
  --- Physics} \textbf{\bibinfo{volume}{2}}, \bibinfo{pages}{419 }
  (\bibinfo{year}{2001}), ISSN \bibinfo{issn}{1296-2147}.

\bibitem[{\citenamefont{Negretti et~al.}(2008)\citenamefont{Negretti, Henkel,
  and M\o{}lmer}}]{Negretti}
\bibinfo{author}{\bibfnamefont{A.}~\bibnamefont{Negretti}},
  \bibinfo{author}{\bibfnamefont{C.}~\bibnamefont{Henkel}}, \bibnamefont{and}
  \bibinfo{author}{\bibfnamefont{K.}~\bibnamefont{M\o{}lmer}},
  \bibinfo{journal}{Phys. Rev. A} \textbf{\bibinfo{volume}{78}},
  \bibinfo{pages}{023630} (\bibinfo{year}{2008}).

\bibitem[{\citenamefont{Dziarmaga and Sacha}(2002)}]{Dziarmaga}
\bibinfo{author}{\bibfnamefont{J.}~\bibnamefont{Dziarmaga}} \bibnamefont{and}
  \bibinfo{author}{\bibfnamefont{K.}~\bibnamefont{Sacha}},
  \bibinfo{journal}{Phys. Rev. A} \textbf{\bibinfo{volume}{66}},
  \bibinfo{pages}{043620} (\bibinfo{year}{2002}).

\bibitem[{\citenamefont{Castin}(2001)}]{Castin}
\bibinfo{author}{\bibfnamefont{Y.}~\bibnamefont{Castin}}, in
  \emph{\bibinfo{booktitle}{Les Houches Session LXXII, Coherent Atomic Matter
  Waves 1999}}, edited by
  \bibinfo{editor}{\bibfnamefont{R.}~\bibnamefont{Kaiser}},
  \bibinfo{editor}{\bibfnamefont{C.}~\bibnamefont{Westbrook}},
  \bibnamefont{and} \bibinfo{editor}{\bibfnamefont{F.}~\bibnamefont{David}}
  (\bibinfo{publisher}{Springer-Verlag}, \bibinfo{address}{Berlin, Germany},
  \bibinfo{year}{2001}).

\bibitem[{\citenamefont{Peierls and Yoccoz}(1957)}]{Peierls}
\bibinfo{author}{\bibfnamefont{R.~E.} \bibnamefont{Peierls}} \bibnamefont{and}
  \bibinfo{author}{\bibfnamefont{J.}~\bibnamefont{Yoccoz}},
  \bibinfo{journal}{Proceedings of the Physical Society. Section A}
  \textbf{\bibinfo{volume}{70}}, \bibinfo{pages}{381} (\bibinfo{year}{1957}).

\bibitem[{\citenamefont{Madsen and M\o{}lmer}(2001{\natexlab{b}})}]{Larsrot}
\bibinfo{author}{\bibfnamefont{L.~B.} \bibnamefont{Madsen}} \bibnamefont{and}
  \bibinfo{author}{\bibfnamefont{K.}~\bibnamefont{M\o{}lmer}},
  \bibinfo{journal}{Phys. Rev. A} \textbf{\bibinfo{volume}{64}},
  \bibinfo{pages}{060501} (\bibinfo{year}{2001}{\natexlab{b}}).

\bibitem[{\citenamefont{Yarkony}(1996)}]{RevModPhys.68.985}
\bibinfo{author}{\bibfnamefont{D.~R.} \bibnamefont{Yarkony}},
  \bibinfo{journal}{Rev. Mod. Phys.} \textbf{\bibinfo{volume}{68}},
  \bibinfo{pages}{985} (\bibinfo{year}{1996}).

\bibitem[{\citenamefont{Zhang and Dong}(2010)}]{ZangDong}
\bibinfo{author}{\bibfnamefont{J.~M.} \bibnamefont{Zhang}} \bibnamefont{and}
  \bibinfo{author}{\bibfnamefont{R.~X.} \bibnamefont{Dong}},
  \bibinfo{journal}{European Journal of Physics} \textbf{\bibinfo{volume}{31}},
  \bibinfo{pages}{591} (\bibinfo{year}{2010}).

\bibitem[{\citenamefont{Liang}(1995)}]{ShoudanLiang}
\bibinfo{author}{\bibfnamefont{S.}~\bibnamefont{Liang}},
  \bibinfo{journal}{Comp. Phys. Commun.} \textbf{\bibinfo{volume}{92}},
  \bibinfo{pages}{11} (\bibinfo{year}{1995}), ISSN \bibinfo{issn}{0010-4655}.

\end{thebibliography}

\end{document}